\documentclass[equations,12pt]{article}
\usepackage{graphicx,epsfig}
\input epsf
\def\LAMBDABAR {\hbox{$\lambda$\kern-0.52em\raise+0.45ex\hbox{--}\kern+0.2em}}
\def\rsim{>\kern-2.5ex\lower0.85ex\hbox{$\sim$}\ }
\def\lsim{<\kern-2.5ex\lower0.85ex\hbox{$\sim$}\ }

\def\ni{\noindent}

\begin{document}

\baselineskip 18pt

 \centerline{\large\bf Search for higher dimensions through their}
\centerline {\large\bf gravitational effects in high energy
collisions}

 \vspace{.25in}

\centerline{Adrian C. Melissinos}

\centerline{Department of Physics and Astronomy, University of
Rochester} \centerline{ Rochester NY 14627 }

\centerline{April 18, 2006}

\vspace{.50in}

\begin{abstract}We consider the use of a microwave parametric
converter for the direct detection of gravitational effects at the
LHC. Because of the extra dimensions the strength of the
gravitational interaction in the bulk grows at high energies. This
leads to possibly detectable signals.
\end{abstract}

The existence of additional dimensions has been widely discussed
in connection with string theory [1].  It has been proposed that
the extra dimensions would modify the gravitational interaction so
that it becomes strong at an energy scale well below the Planck
mass, possibly as low as a few TeV.  Experimental signatures at
the Large Hadron Collider (LHC) have been identified [2,3].  Here
we speculate on the direct detection of the gravitational effects
in high energy particle collisions by using an electromagnetic
detector which couples directly to deformations of the space-time
(3+1) metric [4].

We consider two cases and in both instances we exploit the
periodic nature of the signal which is related to the collision
frequency at the LHC, $f = 31.6$ MHz.  In the first case we
examine the deformation of the metric due to the passage of the
particle bunch in the vicinity of the detector.  This is a
kinematic rather than a dynamic effect.  In the second case we
assume that during the collision, gravitational radiation is
emitted into the higher dimensions (the bulk).  We argue that as a
consequence an \lq\lq evanescent" gravitational wave is generated
on the brane, and it is this disturbance of the metric that is
detected [5]

\vspace{.250in}

\ni {\bf 1. Passage of a relativistic particle}

When a fast-moving particle of mass $m$ passes by a stationary
observer the metric is deformed and at closest approach,  at a
distance $b$, is

\begin{equation}
h_{oo} = \frac{ 2mG}{c^2b} \left( 2\gamma^2\beta^2 + 1
\frac{}{}\right)
\end{equation}

\ni For details see [6].  The particle also imparts an attractive
impulse on the observer (of mass $M$).  In a circular accelerator
the repeated passage of a bunch of $N$ particles results in an
average acceleration

\begin{equation}
<F/M> = \frac{1}{\tau_0} \int^\infty_{-\infty} \frac{F(t)}{M} dt =
\frac{1}{\tau_0} \frac{2NmG}{\gamma\beta cb} (2\gamma^2\beta^2 +1)
\end{equation}

\ni Here $\tau_0$ is the revolution period of the bunch and as
before, $b$ is the distance of closest approach between the bunch
and the observer.

The proposed detector is a microwave parametric converter using
superconducting cavities [4,7,8]. The carrier frequency in the
detector is in the GHz range, and the detection frequency can be
selected at will.  Such a detector can detect metric deformations
of order $h_{\mu\nu} \sim 10^{-23}$ in an integrating time $\tau =
10^4$ s.  Using the LHC parameters

$$\begin{array}{lll}
N/{\rm bunch} & = & 1.15 \times 10^{11}\\
\gamma & =& 7.461 \times 10^3
\end{array}$$

\ni and $b = 0.1$ m, Eq(1)  yields $h_{oo} = 2 \times 10^{-38}$,
well below the detector sensitivity.

Could this result be modified by the presence of extra dimensions?
We recall that the energy scale, $M_G$, at which gravitation
becomes strong is related to the Planck scale, $M_P$ through [3]

\begin{equation}
M_G^{D+2} = M_P^2\left(\frac{\hbar}{Lc}\right)^D
\end{equation}

\ni Here $D$ is the number of extra dimensions and $L$ is the
compactification distance.  In this regime, that is at high
energies and for distances $b<L$, one should replace the Newtonian
constant, $G$, in Eq.(1) by $G(M_P/M_G)^2$.   Thus the signal
would be detectable if $M_G \sim 10^{11}$ GeV.  For this value of
$M_G$ the compactification distance, as given by Eq.(3), is

\begin{eqnarray}
M_G = 10^{11}\ {\rm GeV}\qquad  \qquad D & = 1  & \qquad  L = 2
\times 10^{-9}\
{\rm cm}\nonumber \\
 D & =  2 & \qquad  L = 2 \times 10^{-17}\ {\rm cm}
 \end{eqnarray}

\ni These distances are much smaller than $b$, thus invalidating
the above argument.

One can consider instead the currently most \lq\lq popular" value
for $M_G \sim 10^3$ GeV, in which case

\begin{eqnarray}
M_G = 10^3\ {\rm GeV} \qquad \qquad  D & = 1 &\qquad L = 2 \times
10^{15}\ {\rm
cm}\nonumber \\
 D & = 2 & \qquad L = 0.2\ {\rm cm}
\end{eqnarray}

\ni The first of these values is excluded from experimental
evidence on the validity of Newton's law at large distances.  For
$D=2$ the compactification distance is shorter than the proposed
distance to the detector, $b=10$ cm.  However because of the
highly relativistic motion we can argue that the (transverse ?)
distance $b$ is foreshortened by a factor of $\gamma$ or, as seen
by the observer, the bunch passes by at a distance $\overline{b} =
b/\gamma \sim 10^{-3}$ cm.

Finally we mention that a similar experiment carried out at
Fermilab using 800 GeV protons observed a signal which the authors
attributed to electromagnetic background [9].  If  that signal was
to be interpreted in the formalism presented above it would
correspond to $M_G \sim 10^9$ GeV, and a $D=1$ compactification
distance $L=2 \times 10^{-3}$ cm.

\vspace{.25in}

\ni {\bf 2. Graviton emission}

In this case we consider the emission of gravitons into the bulk.
The cross-section depends on the energy scale where gravity
becomes strong and for energetic gravitons varies in the range
$1-10^4$ fb. [9] Assuming the upper limit and luminosity
$L=10^{34}$ cm$^{-2}$ s$^{-1}$ results in a disappointingly low
rate for a resonant detector, such as we propose to use.

On the other hand, we are interested in the evanescent wave that
propagates on the brane and that extends only a few wavelengths
into the \lq\lq wave zone".  We argue that this wave should be
present for every collision and there are $\sim$ 20 collisions per
crossing.  The frequency seen by the detector is determined by the
time interval between collisions and thus fixed at $f = 36.1$ MHz.

A difficulty with the detection of high frequency gravitational
waves is that for a detectable amplitude $(h_{\mu\nu})$ the energy
density in the wave is very high (grows as $f^2$).  This issue is
not present in the case of the evanescent wave since it does not
carry energy away from the interaction.  The energy density in the
wave that propagates in the bulk is much lower because $G$ has
become strong in the extra dimensions.  Thus the proposed process
is not energetically forbidden.  Clearly, a numerical estimate of
$h_{\mu\nu}$ is necessary in order to evaluate the feasibility of
this proposal.

Here we have not addressed the problem of electromagnetic coupling
to the detector and of other backgrounds.  In principle such
issues are manageable if a detectable signal is present.

\vspace{.50in}

\ni {\bf References}

\begin{description}

\item{[1]} N. Arkani-Hamed, S. Dimopoulos and G. Dvali, Phys. Rev.
{\bf D59} 086004 (1999)

\item{[2]} I. Antoniadis et al., Phys. Lett. {\bf B460}, 176
(1999)

\item{[3]} P. Binetruy et al. Phys. Lett. {\bf B477}, 285 (2000)

\item{[4]} F. Pegoraro et al., Phys. Lett. {\bf A68}, 165 (1978)

\item{[5]} I thank Iosif Bena for a discussion of this issue.

\item{[6]} A.C. Melissinos, Nuovo Cimento, {\bf 62B}, 190 (1980)

\item{[7]} C.E. Reece et al., Nuclear Instruments and Methods,
{\bf A245}, 299 (1986)

\item{[8]} R. Ballantini et al., \lq\lq Microwave apparatus for
gravitational wave detection", ArXiv/gr-qc/0502054

\item{[9]} P. Reiner et al., Phys. Lett. {\bf176}, 233 (1986)

\item{[10]} G.F. Guidice et al., Nucl. Phys. {\bf B544}, 3 (1999)

\end{description}

\end{document}